\documentclass[prl,twocolumn,showpacs,floatfix,superscriptaddress]{revtex4}
\usepackage{epsfig}
\usepackage{amsmath,amssymb}
\usepackage{graphicx}
\usepackage{dcolumn}
\usepackage{bm}
\begin{document}
\title{Efficient Photoionization-Loading of \\
            Trapped Cadmium Ions with Ultrafast Pulses}
\author{L. Deslauriers}
 \affiliation{Department of Physics, Stanford University, Stanford, CA  94305}
\author{M. Acton}
\affiliation{FOCUS Center, Optical Physics Interdisciplinary Laboratory, 
and Department of Physics, University of Michigan, Ann Arbor, MI  48109}
\author{B. B. Blinov}
 \affiliation{Department of Physics, University of Washington, Seattle, WA  98195}
\author{K.-A. Brickman}
\affiliation{FOCUS Center, Optical Physics Interdisciplinary Laboratory, 
and Department of Physics, University of Michigan, Ann Arbor, MI  48109}
\author{P. C. Haljan}
 \affiliation{Physics Department, Simon Fraser University, Burnaby, BC V5A 1S6, Canada}
\author{W. K. Hensinger}
 \affiliation{Department of Physics and Astronomy, University of Sussex, Brighton BN1 9RH, UK}
\author{D. Hucul}
\affiliation{FOCUS Center, Optical Physics Interdisciplinary Laboratory, 
and Department of Physics, University of Michigan, Ann Arbor, MI  48109}
\author{S. Katnik}
\affiliation{FOCUS Center, Optical Physics Interdisciplinary Laboratory, 
and Department of Physics, University of Michigan, Ann Arbor, MI  48109}
\author{R. N. Kohn, Jr.}
\affiliation{FOCUS Center, Optical Physics Interdisciplinary Laboratory, 
and Department of Physics, University of Michigan, Ann Arbor, MI  48109}
\author{P. J. Lee}
 \affiliation{National Institute of Standards and Technology, Atomic Physics Division, Gaithersburg, MD  20899}
\author{M. J. Madsen}
 \affiliation{Department of Physics, Wabash College, Crawfordsville, IN  47933}
\author{P. Maunz}
\affiliation{FOCUS Center, Optical Physics Interdisciplinary Laboratory, 
and Department of Physics, University of Michigan, Ann Arbor, MI  48109}
\author{S. Olmschenk}
\affiliation{FOCUS Center, Optical Physics Interdisciplinary Laboratory, 
and Department of Physics, University of Michigan, Ann Arbor, MI  48109}
\author{D. L. Moehring}
\affiliation{FOCUS Center, Optical Physics Interdisciplinary Laboratory, 
and Department of Physics, University of Michigan, Ann Arbor, MI  48109}
\author{D. Stick}
\affiliation{FOCUS Center, Optical Physics Interdisciplinary Laboratory, 
and Department of Physics, University of Michigan, Ann Arbor, MI  48109}
\author{J. Sterk}
\affiliation{FOCUS Center, Optical Physics Interdisciplinary Laboratory, 
and Department of Physics, University of Michigan, Ann Arbor, MI  48109}
\author{M. Yeo}
\affiliation{FOCUS Center, Optical Physics Interdisciplinary Laboratory, 
and Department of Physics, University of Michigan, Ann Arbor, MI  48109}
\author{K. C. Younge}
\affiliation{FOCUS Center, Optical Physics Interdisciplinary Laboratory, 
and Department of Physics, University of Michigan, Ann Arbor, MI  48109}
\author{C. Monroe}

\email{crmonroe@umich.edu}
\affiliation{FOCUS Center, Optical Physics Interdisciplinary Laboratory, 
and Department of Physics, University of Michigan, Ann Arbor, MI  48109}
\date{\today}
%\preprint{Ver 2}

\begin{abstract}
Atomic cadmium ions are loaded into radiofrequency ion traps by photoionization of
atoms in a cadmium vapor with ultrafast laser pulses. The photoionization is driven through an intermediate atomic resonance with a frequency-quadrupled mode-locked Ti:Sapphire laser that produces pulses of either $100$ fsec or $1$ psec duration at a central wavelength of $229$ nm. The large bandwidth of the pulses photoionizes all velocity classes of the Cd vapor, resulting in high loading efficiencies compared to previous ion trap loading techniques. Measured loading rates are compared with a simple theoretical model, and we conclude that this technique can potentially ionize every atom traversing the laser beam within the trapping volume.  This may allow the operation of ion traps with lower levels of background pressures and less trap electrode surface contamination.  The technique and laser system reported here should be applicable to loading most laser-cooled ion species. 
\end{abstract}
\maketitle

\section {Introduction}

Electromagnetic ion traps have become a standard tool in many areas of physics, 
from mass spectroscopy \cite{Paul,mass} and precision frequency metrology \cite{clocks,Dehmelt} to quantum information science and fundamental studies of quantum mechanics \cite{QIS, phystoday, steaneQC,  NISTJR}. An important practical aspect of operating an ion trap is the efficient and controlled loading of ions into the trap. The standard method for producing positively charged trapped ions is electron bombardment of neutral atoms, usually from electron beams having energies typically of order 100 eV, comfortably above typical ionization thresholds. While this method applies to any atomic or molecular species, electron-beam loading of ions can adversely impact ion trap performance by degrading vacuum quality, charging nearby insulators, and corrupting the surface quality of the trap electrodes \cite{devoe}.  For applications such as atomic clocks or quantum computing, where a specific ion species is to be loaded, photoionization-loading is an attractive alternative that can be much more efficient and cleaner than electron bombardment loading \cite{Gulde01}.  

Several groups have enjoyed the benefits of narrowband photoionization-loading with particular
atomic ions, including Mg$^+$ \cite{Kjaergaard00, Madsen00}, Ca$^+$ \cite{Gulde01, Lucas04}, Yb$^+$ \cite{YbNPL,YbPI} and Sr$^+$ \cite{Sr}. Loading of ions is fairly efficient in these systems, with the continuous-wave (cw) photoionizing lasers tuned to an intermediate atomic resonance en route to ionization.  Furthermore, the use of narrowband photoionizing lasers allows isotopic selectivity, as the optical isotope shift in the intermediate resonance is typically larger than the resonant linewidth of the atom.  

Here, we report the use of ultrafast laser pulses to photoionize and load cadmium ions (Cd$^+$) in a variety of rf trap geometries, with even greater efficiency.  The laser pulses are tuned to an intermediate resonance in the neutral Cd atom, and the same pulses (or a cw laser beam used for laser-cooling the eventual ion) can then directly promote the electron to the continuum. The large bandwidth of the pulsed laser can ionize all velocity classes of atoms in a vapor or atomic beam and can provide nearly perfect ion trap loading efficiency. As a result, the density of atoms at the trapping region can be reduced to a level where the atomic coating of the electrodes is negligible. This not only permits a potentially lower vacuum level near the trapping region, but may also reduce anomalous heating of trapped ions that could arise from contaminated electrodes \cite{NISTJR,Turchette00,devoe,Louis04}. 

\section {Photoionization scheme and experimental setup}
\begin{figure}[hptb]
\centering
\includegraphics[width=\linewidth,clip]{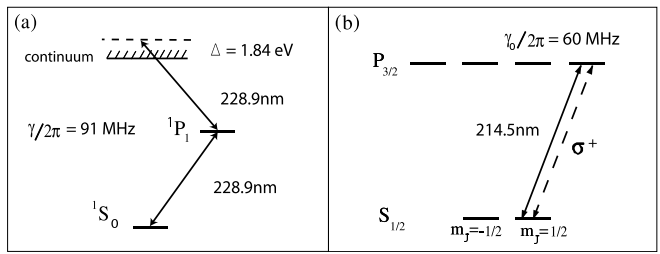}
\caption{(a) Relevant energy levels for the neutral Cd atom. From the two-electron $^1S_0$ ground state, the intermediate excited $5s5p\ ^1P_1$ state (radiative linewidth $\gamma/2\pi=91$ MHz) is populated with a pulsed laser near $\lambda = 228.9$ nm.  This state is well above the midpoint of the energy difference between the $5s^2\ ^{1}S_{0}$ ground state and the continuum, so the same laser can subsequently ionize the atom.   (b) Relevant energy levels for even isotopes of the Cd$^+$ ion. A continuous wave laser tuned red of the $5s$  $^{2}S_{1/2}$ $\rightarrow$ $5p$ $^{2}P_{3/2}$ cycling transition near $\lambda = 214.5$ nm (radiative linewidth of $\gamma_{o}/2\pi = 60$ MHz) provides Doppler cooling of the motion for the newly formed ions and localizes them to the center of the rf trap.  For odd isotopes, there is also hyperfine structure.} 
\label{EnergyLevels}
\end{figure}

The relevant energy levels of the neutral Cd atom are shown in Fig. \ref{EnergyLevels}(a). The output of a quadrupled mode-locked Ti-Sapphire (Ti:S) laser at a wavelength of $228.9$ nm is resonant with the $5s^{2}$ $^1S_0$ $\rightarrow 5s5p$ $^1P_1$ transition in Cd (radiative linewidth $\gamma/2\pi = 91$ MHz). The same laser then promotes the intermediate $P$ state population to the continuum, $1.84$ eV above the ionization threshold. Fig. \ref{EnergyLevels}b shows the energy levels of the singly ionized Cd atom. Typically during the ion trap loading process, a narrowband cw laser tuned near resonance of the ${5s\,^{2}S_{1/2}}$ $\rightarrow$ ${5p\,^{2}P_{3/2}}$ transition in the Cd$^+$ ion at a wavelength of $214.5$ nm is also directed into the trap, providing Doppler laser cooling of the newly formed ions. The resulting fluorescence light from the ions is collected and imaged onto a CCD camera in order to detect the presence of ions loaded into the trap. Note that the photons in the cw laser-cooling beam have enough energy to ionize neutral Cd atoms from the intermediate $P$ state.
%--[hptb] means p=seperate page, t=top page,b...,h=here
%-----------------------FIG---ENERGY LEVELS   ----------------------------

Most trapped atomic ion species that are easily laser cooled (those whose neutral form have two valence $S$ electrons: Be, Mg, Ca, Sr, Ba, Zn, Cd, Hg, and Yb) have an intermediate excited $^{1}P_{1}$ state more than halfway to the continuum, and can therefore be ionized using the same two-photon process described here.  The various laser wavelengths required to couple the $^{1}S_{0}$ ground state to a selected $^{1}P_{1}$ excited state in these atomic species are shown in Table \ref{tab:AtomicSpecies}.  In the case of Yb, the intermediate state is just below halfway to the continuum, but in practice ionization is still effective with a single laser coupling to highly-lying Rydberg states that rapidly ionize in the trap electric fields \cite{YbPI}.  For Ca, Sr, and Ba, the selected intermediate state is not the lowest $^1P_1$ level. 
%We note that for Sr and Ba, it should be possible to ionize the atom directly from the $^{1}S_{0}$ ground state using a reasonable ultraviolet laser.   
%--------------------------------------------TABLE--WAVELENGTHS------------
\begin{table}
\centering
\caption{Typical atomic species used in laser-cooled ion trapping experiments, with the relevant neutral-atom transition wavelengths from the ground $^{1}S_{0}$ state to the $^{1}P_{1}$ state and to the single ionization threshold. For each atom, the chosen intermediate excited $^{1}P_{1}$ state is near or above the midpoint in the energy separating the ground state and the continuum, making a 2-photon ionizing process feasible with a single laser. (Wavelengths for Ca, Sr, and Ba are not the lowest $^{1}P_{1}$ state.)  The laser setup used in the Cd experiment reported here can be applied to all atomic species listed in the table.\\} 
\space
\begin{tabular}{c|c c}  \hline \hline
Atomic&$\lambda(^{1}S_{0}\rightarrow ^1P_{1})$ &$\lambda(^{1}S_{0}\rightarrow\infty)$ \\
Species & (nm) & (nm) \\ \hline
Be & 234.9 & 133 \\ 
Mg & 285.3 & 162 \\ 
Ca & 272.2 & 203 \\ 
Sr & 293.0 & 218 \\ 
Ba & 350.2 & 238 \\ 
Zn & 213.9 & 132 \\ 
Cd & 228.9 & 138 \\ 
Hg & 185.0 & 119 \\ 
Yb & 398.9 & 198 \\ \hline \hline
\end{tabular}
\label{tab:AtomicSpecies}
\end{table}

\begin{figure*}[hptb]
\centering
\includegraphics[width=\linewidth,keepaspectratio]{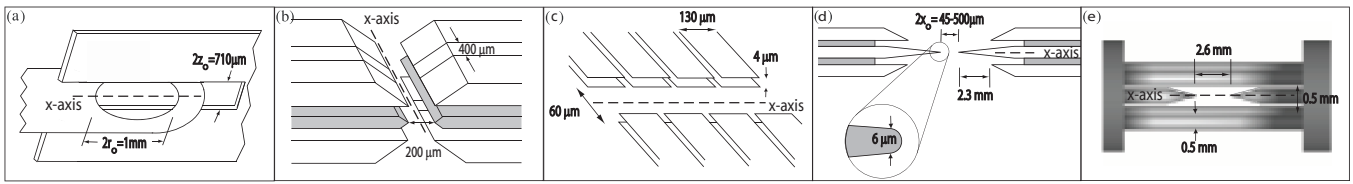}
\caption{Ion traps used in this experiment: (a) ring/fork quadrupole trap \cite{Louis04}, (b) three-layer linear trap \cite{Louis04}, (c) GaAs chip linear trap \cite{Stick06}, (d) two-needle quadrupole trap \cite{LouisNeedle06}, and (e) four-rod linear trap \cite{moehring06}.}
\label{TrapPix}
\end{figure*}

The absolute rate of loading ions in a trap depends on the particular geometry of the ion trap and the flux of neutral atoms entering the trap.  We report photoionization-loading of Cd ions in a number of geometries, including an asymmetric quadrupole ``ring/fork" trap \cite{Jefferts95,moehring06}, a three-layer linear trap \cite{Louis04,Hensinger06}, a microfabricated GaAs linear trap \cite{Stick06}, a symmetric two-needle quadrupole trap \cite{LouisNeedle06}, and a large four-rod linear trap \cite{moehring06}. In Fig. \ref{TrapPix}, each ion trap is drawn along with corresponding dimensions. Relevant characteristics of the traps, such as trap depth, loading volume, ion oscillation and rf drive frequencies are listed in Table \ref{tab:IonTraps}. The trapping volumes span a wide range, from a linear dimension of $45\,\mu$m for the two-needle trap up to almost $1$ mm for the four-rod trap. The trap depths range from below $0.1$ eV in the case of the GaAs microfabricated trap and the double-needle trap to several eV for the larger traps. 

%---------------------------------FIG---TRAP PICTURES---------------------------
%---------------------------------------------------------------------------
%-----------------------------------TABLE--TRAP PARAMETERS-----------------------
\begin{table*}
\centering
\caption{Characteristics of linear and quadrupole ion traps of Fig.~\ref{TrapPix} used for ultrafast photoionization loading.  The trap size $L$ is defined as the approximate linear dimension of the trapping volume along the photoionizing beam.  The ion secular oscillation frequencies $\nu_x,\,\nu_y,$ and $\nu_z$ are taken along the principal axes of the trap, with the $x$-axis denoted in Fig. \ref{TrapPix}. \\}
\begin{tabular}{c||c|c|c|c|c}  \hline \hline
\small{Trap Description} & \small{ring/fork quadrupole} & \small{three-layer linear} & \small{GaAs chip linear} 
                      & \small{double-needle quadrupole}  & \small{four-rod linear}  \\ \hline
\small{trap depth (eV)}    & \small{0.8}             & \small{0.2-5.0}        & \small{0.08-0.13} 
                      & \small{0.02-5.0}        & \small{0.5-2.0}       \\ 
\small{trap Size L ($\mu$m)} & \small{500} & \small{300} & \small{60} & \small{45-500} & \small{700} \\
\small{rf drive (MHz)}   &\small{50}&\small{47}& \small{16}& \small{29}& \small{36}\\
\small{$\nu_x$ (MHz)} & \small{0.5}      & \small{0.6-4.0}   & \small{0.8-1.0} 
                                         & \small{0.5-10.0}  & \small{0.25-0.70}   \\
\small{$\nu_y$ (MHz)} & \small{0.75}     & \small{8.1}       & \small{3.3} 
                                         & \small{0.25-5.0}   & \small{0.90}        \\
\small{$\nu_z$ (MHz)} & \small{1.25}     & \small{8.3}       & \small{4.3} 
                                         & \small{0.25-5.0}   & \small{0.91}       \\ \hline \hline
\end{tabular}
\label{tab:IonTraps}
\end{table*}
%------------------------------------------------------------------------------
We perform the experiment with one of two ultrafast mode-locked lasers, that are frequency-quadrupled to a central wavelength of $\lambda=228.9$ nm for excitation of the neutral Cd atom (Fig. \ref{EnergyLevels}a).  The first pulsed laser provides excitation at a $1$ picosecond time scale (psec laser) while the second laser provides excitation at a $100$ femtosecond time scale (fsec laser), allowing a study of the photoionization for different pulse durations. 

\begin{figure}[hptb]
%--[hptb] means p=seperate page, t=top page,b...,h=here
\centering
\includegraphics[width=0.8\linewidth,clip]{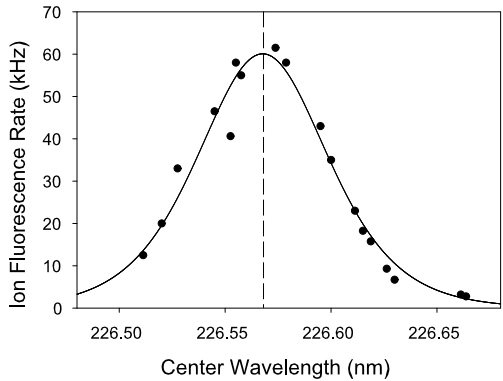}
\caption{Fluorescence spectrum of a single trapped $^{111}$Cd$^+$ ion as the center frequency of the psec laser is scanned \cite{BBcool}.  The observed $0.07$ nm ($400$ GHz) bandwidth is consistent with a transform-limited pulse of duration $\tau \approx 1$ psec.  The dashed line indicates the expected resonance position of the ${^2S_{1/2} \rightarrow ^2P_{1/2}}$ transition in $^{111}$Cd$^+$ at $226.57$ nm.}
\label{psecBW}
\end{figure}

The psec pulses are generated by a commercial mode-locked Ti:S laser (Spectra-Physics Tsunami), producing pulse energies up to $12$ nJ at $915$ nm.  A second-harmonic-generation autocorrelator is used to measure the pulse duration of approximately $2$ ps.  The infrared pulses are frequency-doubled in a 12 mm long lithium borate (LBO) crystal and doubled again in a 10 mm long $\beta-$barium borate (BBO) crystal. Each nonlinear crystal is critically phase-matched (angle-tuned) at its corresponding wavelength. The frequency-quadrupled output at $\lambda=228.9$ nm consists of pulses of approximate duration $\tau\approx 1$ ps with energies up to $60$ pJ in the trapping region. The bandwidth of the psec laser pulses in the ultraviolet is indirectly determined by tuning the same laser to the nearby $^2S_{1/2} - ^2P_{1/2}$ transition in Cd$^+$ and measuring the trapped ion fluorescence rate \cite{BBcool}.  Figure \ref{psecBW} displays the fluorescence level of a single ion from the quadrupled psec laser tuned near this transition at $226.5$ nm (fundamental infrared laser center wavelength at 906 nm). The resulting ultraviolet bandwidth is consistent with the autocorrelator measurements extrapolated from the infrared.  We expect a very similar spectrum when the laser is tuned to  915 nm and quadrupled to the ${^{1}S_{0}\rightarrow^{1}~P_{1}}$ neutral Cd atom resonance at 228.9 nm, verified in Fig. \ref{LoadingRateVsDetuningPsec} below.

%---------------------------------FIG---PSEC ON ION---------------------------
\begin{figure}[hptb]
\centering
\includegraphics[width=0.8\linewidth]{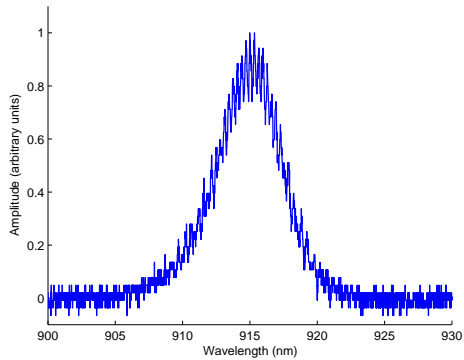}
\caption{Measured spectrum of the fsec laser at a central wavelength of 915 nm. The FWHM bandwidth of approximately $6$ nm is consistent with the production of $\tau\approx 200$ fsec pulses in the infrared.  When quadrupled to the ultraviolet, we expect pulses of duration around $100$ fsec.} 
\label{fsec_spectrum}
\end{figure}

%---------------------------------------------------------------------------
The fsec mode-locked laser is custom built and based largely on the design of Ref. \cite{Asaki04}. Typically, mode-locked Ti:S lasers operate in the optimal region of the gain spectrum, near $800$ nm. While the design and operation of such lasers is well documented, several important issues arise as the wavelength of the pulsed laser is tuned above $900$ nm. A detailed description of the fsec laser cavity and relevant operational issues is provided in Appendix A. The fsec laser typically produces pulse energies of about $6$ nJ at $915$ nm at a repetition rate of $1/T = 86$ MHz.  This light is frequency-doubled in a $7$ mm long LBO crystal and doubled again in a $5$ mm long BBO nonlinear crystal, similar to the psec laser.  The frequency-quadrupled output at $\lambda=228.9$ nm consists of pulses with energies up to $60$ pJ in the trapping region. A spectrum analyzer with a resolution under $1$ nm is used to determine the pulse bandwidth (at 915 nm), as shown in Fig. \ref{fsec_spectrum}.  Given the slight dispersion in the doubling crystals, we expect near transform-limited pulses of approximate duration $\tau \approx 100$ fsec at $228.9$ nm.

%------------------------FIG---FSEC SPECTRUM----------------
%--[hptb] means p=seperate page, t=top page, b...,h=here
\begin{figure}[hptb]
%--[hptb] means p=seperate page, t=top page,b...,h=here\begin{figure}[hptb]
\centering
\includegraphics[width=0.8\linewidth,clip]{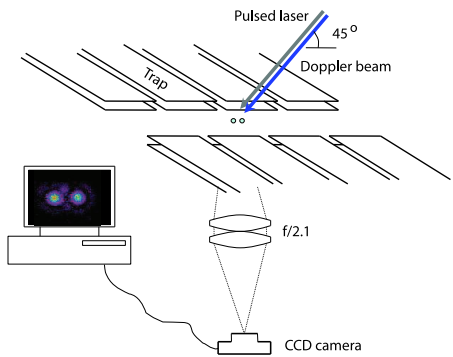}
\caption{Schematic for the layout of the experiment, with the GaAs chip trap of Fig \ref{TrapPix}c.}
\label{ExperimentalLayout}
\end{figure}

%----------------------------------------------------------------------------  
The layout of the experiment is depicted in Fig. \ref{ExperimentalLayout} in the case of the trap in Fig. \ref{TrapPix}c. The cw cooling laser enters the trapping region with a wavevector that has components along all three principle axes of the ion trap, providing effective Doppler cooling in all three dimensions \cite{Itano82}. The scattered light from the ion(s) is captured by an $f/2.1$ microscope objective and sent onto an intensified CCD camera. All of the ion traps are loaded from a room-temperature vapor of cadmium, with an estimated partial pressure of about $10^{-11}$ torr.  This produces a sufficient number of neutrals that traverse the trapping volume for loading, with observed loading rates up to about $10$ ions/sec.  At the same time, this low-level vapor does not compromise the lifetime of the ions in the rf traps through collisions or charge-exchange processes. (We expect that the rate of Langevin elastic collisions with background atoms to be about $10^{-3}$ sec$^-1$ at this pressure \cite{NISTJR}.) The bandwidths of the pulsed lasers used in all of the experiments are much larger than the Doppler broadened linewidth of optical transitions in the atomic vapor.

%--------------FIG 3  EXPERIMENTAL LAYOUT-----------------------------------------
%----------------------------------------------------------------------------
\section {Theoretical ion production rate}
In this section, we calculate the expected photoionization rate for a two-photon resonantly-assisted ionization process. The result highlights the dependence of the loading rate on controlled experimental parameters such as laser pulse energy and duration, laser beam waist, loading volume, and neutral atom density. (For a similar treatment in the case of a directed atomic beam and continuous-wave photoionization, see Ref. \cite{Lucas04}.) We solve the optical Bloch equations for the photoionization rate of a single atom excited to an intermediate resonance following Fig. \ref{EnergyLevels}(a). We assume the pulsed laser peak intensity to be in the weak perturbative regime \cite{CohenTannoudji}, so that no other electronic states are involved in this process.

The laser pulse train consists of individual pulses of approximate duration $\tau = 100$ fsec or $\tau = 1$ psec separated by a pulse period of $T \approx 12$ nsec.  Each pulse is focused down to peak intensity $I$ at the position of the atom to be ionized, resonantly coupling the initial ground $5s^{2}$ $^1S_0$ electronic state to the intermediate exited ${5s5p\,^1P_1}$ state.  The resulting Rabi rotation angle is $\theta = g\tau$, where the resonant Rabi frequency is ${g = \gamma\sqrt{I/2I_\text{sat}} \gg \gamma}$ and ${I_\text{sat}}$ is the ${^1S_0-^1P_1}$ saturation intensity.  The same laser pulse is capable of sequentially ionizing the atom with photoionization rate from the ${^1P_1}$ state given by $\Gamma =I\sigma/\hbar\omega$, where $\sigma$ is the ${^1P_1}$ photoionization cross-section and $\omega$ is the laser frequency.  

A cw laser with intensity $I_\text{cw}$ used for Doppler laser-cooling of the eventual ion can 
also ionize the neutral atom once in the $^1P_1$ state, with photoionization rate given by ${\Gamma_\text{cw} =I_\text{cw}\sigma_\text{cw}/\hbar\omega_\text{cw} \ll \gamma}$, where the cw laser parameters are defined analogously to that of the pulsed laser.

The $^1P_1$ spontaneous emission lifetime is usually much shorter than the period between laser pulses $(T \gg 1/\gamma \gg \tau$), so the neutral atom returns to the ground $^1S_0$ state before the next pulse (unless it has ionized). By integrating the optical Bloch equations for the atomic populations in time (see Appendix B), we can therefore write down the probability $P_\text{ion}$ of ionization per laser pulse period $T$:
\begin{equation}
P_\text{ion} =1 - e^{-\Gamma\tau\!/2}\!\left[1+\frac{\Gamma\tau}{2}\frac{\sin\theta}{\theta}
+ \frac{\Gamma_\text{cw}}{2\gamma}(1\!-\!\cos\theta) \right]\, .
\label{ionprob}
\end{equation}
In this expression, we have assumed that the photoionization rate from the $^1P_1$ state is much weaker than the coherent coupling between the $^1S_0$ and $^1P_1$ states in the atom, or $\Gamma \ll g$.  This is valid for typical atomic systems considered here and for typical (perturbative) laser intensities.

All but the last term in Eq. \ref{ionprob} arise from photoionization from the pulsed laser exclusively, while the last term describes the contribution from the cw laser. For similar average intensities of the cw and pulsed lasers ($I_\text{cw} \approx \overline{I} = I\tau/T$) and assuming that the photoionization cross section from the $^1P_1$ state is similar for the two processes, we find that photoionization probability from the pulsed laser alone is roughly $\gamma T\!/2$ times larger than the photoionization from the cw laser.  In the Cd system considered here ($1/\gamma \approx 2$ nsec and a $1/T \approx 80$ MHz laser repetition rate), this amounts to the pulsed laser ionizing the atom about $3$ times the rate of the cw laser.  In the experiments, the cw laser has a lower average intensity than the pulsed laser, so we neglect the effect of the cw laser on the photoionization calculations for simplicity.

For $\Gamma\tau/2 \ll 1$, the ionization probability per pulse from the pulsed laser alone simplifies to
\begin{equation}
P_\text{ion}  =  \frac{\Gamma\tau}{2}\left(1-\frac{\sin\theta}{\theta}\right) 
         \simeq  \frac{\theta^2\Gamma\tau}{16} \, .
\label{IonizationProb}
\end{equation}
The above approximation is accurate to within $25$\% for all Rabi angles $0\le\theta\le\pi$.  In terms of the laser pulse energy $\mathcal{E}$ and duration $\tau$, the ionization probability per pulse for an atom located at the focus of the laser beam (with waist $\rho$) is
\begin{equation}
P^{(0)}_\text{ion} = \frac{\sigma\gamma^2 \mathcal{E}^2 \tau}{8\pi^2 \hbar\omega I_\text{sat} \rho^4} \, .
\label{IonizationProb2}
\end{equation}

Since the ionization cross section from the Cd $^1P_1$ state is not known precisely, we use previous measurements in atomic magnesium \cite{Madsen00}, calcium \cite{Gulde01} and barium \cite{He91} as guides, because their electronic structure is comparable to Cd.  These experiments reported photoionization cross sections from their respective $^1P_1$ states using laser radiation well above the ionization threshold by similar amounts of energy to that considered here. We estimate to an order of magnitude that the Cd $^1P_1$ photoionization cross section is $\sigma \approx 10^{-16}$ cm$^2$ for light at $\lambda=228.9$~nm.
In terms of the laser parameters, we find that the ionization probability per pulse for an atom at the focus of the laser beam is $P^{(0)}_\text{ion} \simeq 0.004 \mathcal{E}^2 \tau/\rho^4$, with $\mathcal{E}$ in pJ, $\tau$ in ps and $\rho$ in $\mu$m.

%----------------------------FIG---FLYING ATOMS------------------------
%--[hptb] means p=seperate page, t=top page, b...,h=here
\begin{figure}[hptb]
\centering
\includegraphics[width=0.7\linewidth,clip]{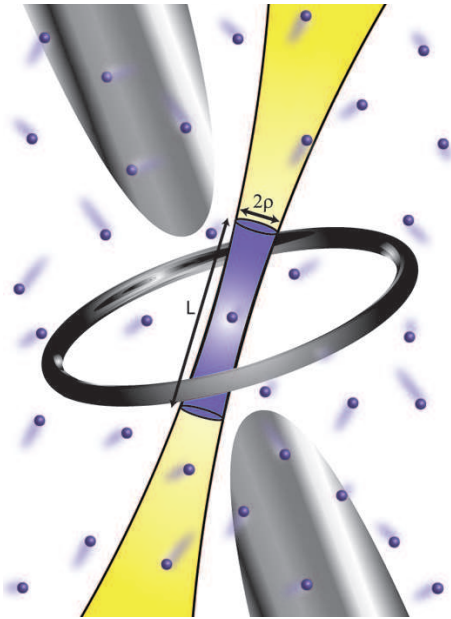}
\caption{Schematic of photoionization laser focused through ion trap volume.  The laser has waist $\rho$ ($1/e$ electric field radius) and overlaps with the ion trap volume over an axial distance $L$.  This gives a cylindrical ion trap loading volume of $\pi \rho^2 L$, assuming the Rayleigh range $\pi\rho^2/\lambda$ of the laser beam to be much larger than $L$.} 
\label{FlyingAtoms}
\end{figure}
%--------------------------------------------------------------------- 
Now we can estimate the total rate of ions produced in a loading volume formed by the intersection of the trapping region and the photoionizing laser beam, depicted by the cylinder of radius $\rho$ and length $L \gg \rho$ in Fig. \ref{FlyingAtoms}.  Here $L$ is defined as the linear dimension of the trapping volume along the axis of the laser beam and is assumed to be much smaller than the Rayleigh range $\pi\rho^2/\lambda$ of the laser beam.  We assume that each atom experiences many laser pulses during its trajectory through the loading volume (for room-temperature thermal Cd atoms with mean speed $\bar{v}=240$ m/s, a typical atom travels about $3\,\mu$m between laser pulses separated by $T \simeq 12$ ns). In Appendix C, the net flux $\Phi_\text{ion}$ of photoionized atoms is calculated by averaging over the thermal trajectories of background atoms in the vapor as they traverse the laser pulses.  In terms of the background density $n_0$, we find
\begin{equation}
\Phi_\text{ion} = \frac{n_0\rho}{8T}P^{(0)}_\text{ion}
= \frac{\sigma\gamma^2 n_0 \mathcal{E}^2\tau}
                       {64\pi^2\hbar\omega I_\text{sat} \rho^3 T} \, .
\label{Flux}
\end{equation}
The average photoionization rate over the entire cylindrical loading volume is then
\begin{eqnarray}
R_\text{ion} &=& \Phi_\text{ion}(2\pi\rho L)
             = \frac{\pi n_0\rho^2 L}{4T}P^{(0)}_\text{ion} \nonumber \\
             &=& \frac{\sigma\gamma^2 n_0 L \mathcal{E}^2\tau}{32\pi\hbar\omega I_\text{sat} \rho^2 T}\,.
\label{IonizationRate}
\end{eqnarray}
In Ref. \cite{Lucas04}, similar expressions are derived for the case of a directed atomic beam and continuous-wave photoionization.
The above treatment assumes that the ionization probability per pulse for a typical atom is small.  If this is not the case, nearly every atom that crosses the laser beam is ionized, and the flux of ionized atoms is then simply $\Phi_\text{ion}=n_0\bar{v}/4$ \cite{Reif}. 

The vapor pressure of Cd metal at room temperature is roughly $10^{-11}$ torr \cite{MotherRussia03, AIPbook}, giving a density of $n_0 = 3 \times 10^5$ atoms/cm$^3$, and we find for the Cd system,
${R_\text{ion} \simeq 1000\,\mathcal{E}^2 (L/\rho^2)(\tau/T)}$, with the pulse energy $\mathcal{E}$ in pJ and the dimensions $L$ and $\rho$ in $\mu$m.  For a $1/T = 80$ MHz laser repetition rate with waist $\rho = 25\,\mu$m traversing a length $L=100\,\mu$m through the trapping volume, we find $R_\text{ion} \simeq 0.01\tau\mathcal{E}^2$ sec$^{-1}$, with $\mathcal{E}$ in pJ and $\tau$ in psec.  For $60$ pJ pulses, we expect a photoionization rate of approximately $R_\text{ion}\simeq 40$ sec$^{-1}$ for the psec laser and $R_\text{ion}\simeq 4$ sec$^{-1}$ for the fsec laser.

The above flux of photoionized atoms can be compared to the total flux of atoms traversing the same region, and the loading efficiency is then
\begin{eqnarray}
\eta &\equiv& \frac{\Phi_\text{ion}}{\Phi_0} 
            = P^{(0)}_\text{ion}\frac{\rho}{2\bar{v}T} \nonumber \\
           &=& \frac{\sigma\gamma^2 \mathcal{E}^2 \tau}
                    {16\pi^2 \hbar\omega I_\text{sat} \bar{v}T\rho^3} \, .
\label{Efficiency}
\end{eqnarray}
For a photoionizing laser repetition rate of $1/T = 80$ MHz, $\eta \simeq 0.001 \mathcal{E}^2\tau/\rho^3$, with $\mathcal{E}$ in pJ, $\tau$ in ps and $\rho$ in $\mu$m.  For $\tau=1$ psec pulses at $\mathcal{E}=60$ pJ focused down to $\rho=10\,\mu$m, the efficiency is expected to be $\eta \simeq 0.4\%$, several orders of magnitude larger than electron bombardment ionization efficiency \cite{Gulde01}.  By increasing the pulse energy to ${\mathcal{E} = 300}$~pJ focused to a waist of $\rho=10\,\mu$m, the efficiency of psec photoionization is expected to approach unity ($P_\text{ion} \sim 1$): nearly every atom that traverses the photoionization laser beam within the trap volume should be ionized.

\section {Experimental Results}
The loading and successful localization of ions in an ion trap through laser cooling is a complex process.  Initially, ions are expected to have high kinetic energies of order the trap depth (up to several eV), and laser cooling can follow a very slow dynamical time scale (seconds) \cite{LaserCool}.  Nonlinearities in the trapping potential, along with the large rf potentials near the edge of the trap, can lead to dynamical instabilities and prevent loading of such energetic ions \cite{Chaos}.  In addition, successful loading relies upon adequate overlap between the photoionization beam and the trapping volume. We do not consider any of these effects when comparing experimental and theoretical photoionization loading rates \cite{electrode}.  The presence of ions already in the trap can also influence the loading process, so for most measurements of loading rate presented below, we simply measure the mean time to load a single ion in a previously empty trap.  

\begin{figure}[hptb]
\centering
\includegraphics[width=0.8\linewidth,clip]{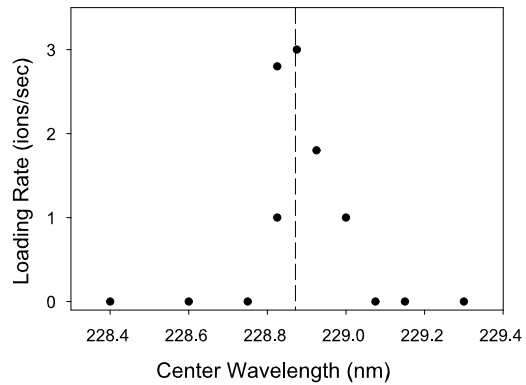}
\caption{Loading rate of Cd$^+$ ions in the ring/fork quadrupole trap of Fig. \ref{TrapPix}a vs. detuning of the psec photoionization laser from the neutral Cd $^1S_0 \rightarrow ^1P_1$ transition at $228.9$ nm. The laser repetition rate is $80$ MHz, and each pulse has approximate duration $1$ ps and energy $60$ pJ. The loading rate tracks the spectrum of the psec laser, with a bandwidth of approximately $400$ GHz.  The dashed line indicates the expected resonance position of the $^1S_0 \rightarrow ^1P_1$ transition in neutral Cd at $228.87$ nm.} 
\label{LoadingRateVsDetuningPsec}
\end{figure}

In Fig. \ref{LoadingRateVsDetuningPsec}, we plot the observed loading rate of ions in the quadrupole trap of Fig \ref{TrapPix}a versus the central wavelength of the psec laser (with approximate values $\mathcal{E} = 60$ pJ, $\tau = 1$ ps, $1/T = 80$ MHz, and $\rho \simeq 25\,\mu$m). The photoionization rate traces the spectrum of the psec laser, with peak loading rate near resonance at a central wavelength of 228.9 nm.  The $0.1$ nm ($\sim\!\!~600$ GHz) bandwidth is consistent with the transform limit expected from the measured pulse duration, and allows the excitation and photoionization of atoms in all velocity classes in the background vapor (Doppler width $\bar{v}/\lambda \approx 1$ GHz).  On resonance (corresponding to $\theta \approx \pi$ \cite{Madsen06}), we find the average loading rate for single Cd$^+$ ions is of order ${R_\text{ion} = 1-10\text{ sec}^{-1}}$.

\begin{figure}[hptb]
\centering
\includegraphics[width=0.8\linewidth,clip]{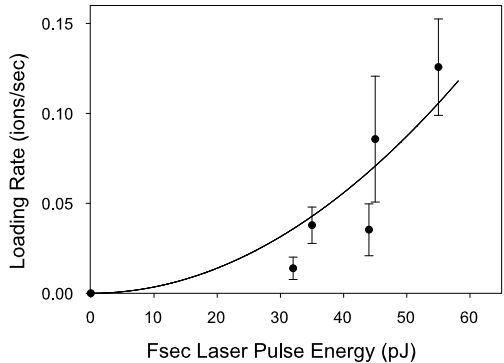}
\caption{Observed loading rate vs. pulse energy in the fsec photoionization laser in the linear trap of Fig. \ref{TrapPix}b.  The pulse energy is varied with attenuators, and the waist of approximately $\rho \approx 25\,\mu$m is held constant.  At the highest energy of $60$ pJ, we estimate that $\theta \simeq 0.3\pi$. The line is the best fit quadratic dependence of loading rate on energy as expected from theory.  The observed absolute loading rate is within an order of magnitude of theory, which is reasonable considering the many other variables that impact successful loading and the uncertainty in the photoionization cross section. The considerable spread in the data may be attributed to varying positions of the pulsed laser within the ion trap volume or varying trap characteristics, as the data was taken over the course of several weeks.} 
\label{LoadingRateVsPowerFsec}
\end{figure}

In Fig \ref{LoadingRateVsPowerFsec}, the observed loading rate in the linear trap of Fig \ref{TrapPix}b is plotted as a function of the average power of the fsec pulsed laser. The observed loading rate is consistent with a quadratic scaling in optical power as predicted for $\theta < \pi$ (Eq. \ref{IonizationRate}).  At the highest pulse energy of about $60$ pJ, the loading rate is about $0.1$ sec$^{-1}$. This is much less than that the psec laser results in Fig. \ref{LoadingRateVsDetuningPsec}, due to the shorter pulse duration $\tau$ and the somewhat smaller trapping size $L$ as predicted from Eq. \ref{IonizationRate}.  While the observed dependence of loading rate on laser parameters is in rough agreement with theory, the absolute loading rates are about an order of magnitude lower than expected, possibly reflecting the inefficient cooling process discussed above, or overestimates of the effective trap dimension $L$ or the photoionization cross section $\sigma$.  

%----------------------------FIG---PSEC DETUNING----------------------------
%--[hptb] means p=seperate page, t=top page, b...,h=here
%--------------------------------------------------------------------- 
%-------------------------------FIG---FSEC POWER--------------------------
%--[hptb] means p=seperate page, t=top page, b...,h=here
%------------------------------------------------------------------------------------ 

In the smallest ion traps with characteristic size ${L\approx 50\,\mu}$m (double-needle and GaAs chip trap of Figs. \ref{TrapPix}c-d)), the loading rate is lower than the larger traps using similar photoionizing laser parameters, as expected from Eq. \ref{IonizationRate}.  For example, in the GaAs chip trap \cite{Stick06}, the loading rate was approximately $0.01$ sec$^{-1}$ with the fsec laser at similar energies to that reported above.

\begin{figure}[hptb]
\centering
\includegraphics[width=0.8\linewidth,clip]{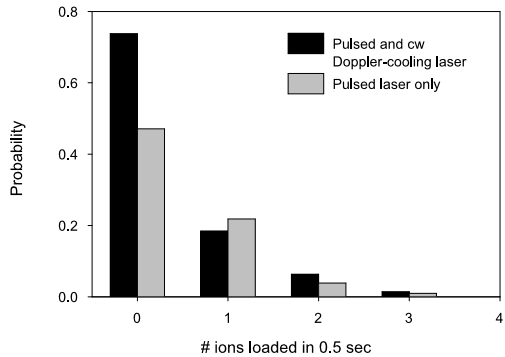}
\caption{Histogram of the number of ions loaded in a $0.5$ sec exposure interval to the fsec pulsed laser in the large 4-rod linear trap of Fig. \ref{TrapPix}e.  The two sets of bars correspond to the presence (206 trials) and absence (152 trials) of the cw Doppler cooling laser beam during the loading process.  The average loading rate is observed to be $0.7$ sec$^{-1}$ ($0.9$ sec$^{-1}$) when the cw Doppler laser beam is present (absent), with the discrepancy not statistically significant.}
\label{histogram}
\end{figure}

In most of the Cd ion traps loaded with pulsed-laser photoionization, the cw Doppler cooling beam has no significant effect on loading: when the pulsed and cw beams are directed into the trap at the same time, the loading rate is observed to be nearly the same as when the lasers are slowly ($<1$ Hz) alternated in sequence.  Fig. \ref{histogram} displays a histogram of the number of ions loaded in a $0.5$ sec exposure time to the fsec photoionization beam in the 4-rod trap (Fig \ref{TrapPix}e) with and without the simultaneous presence of the cw Doppler-cooling beam, yielding similar average loading rates of $R_\text{ion} \simeq 1$ sec$^{-1}$. However, the linear trap of Fig \ref{TrapPix}b behaves markedly different than the others.  In this trap, the presence of the cw Doppler cooling beam during photoionization is observed to enhance the loading rate by up to an order of magnitude. In addition, by tuning the cw laser near resonance in a particular Cd$^+$ isotope in this trap, we are able to selectively load particular isotopes---a feature that is impossible with the pulsed laser alone, as it has a bandwidth much larger than the $1-5$ GHz isotope splittings in Cd.  One possible explanation for this behavior is that the linear trap may require a higher level of initial laser cooling for ions to remain confined, perhaps from an undue level of rf micromotion \cite{Dehmelt}.

%----------------------------FIG---HISTOGRAM--------------
%--[hptb] means p=seperate page, t=top page, b...,h=here
%--------------------------------------------------------------------- 
\section {Conclusion}
We have demonstrated efficient photoionization loading of Cd$^+$ ions into a variety of rf ion traps using frequency-quadrupled Ti:S mode-locked pulsed lasers.  The laser pulses resonantly promote neutral Cd atoms to an intermediate excited state that subsequently photoionizes.  Observed loading rates over $1$ ion/sec are within an order of magnitude of the expected photoionization rate derived from a simple theoretical model of the photoionization process, and the dependence on laser parameters and trap volume agree with theory.  Because the pulses are broadband, all velocity classes of atoms can be photoionized, leading to a very efficient loading technique.  With slightly increased pulse energies and tighter focusing, we expect that ultrafast laser pulses could successfully load every atom that traverses the laser beam within a typical trapping volume.  This may allow ion trap electrodes to be cleaner than in previous systems, with lower pressures in the trapping region.

\appendix
\section{Appendix A: Description of Long-Wavelength Mode-Locked Ti:Sapphire Laser}
The mode-locked Ti:Sapphire (Ti:S) laser is very useful for photoionization-loading of trapped ions because of its large tuning range ($650-1150$ nm).  Moreover, with ultrafast pulses in the range $10-100$ fsec, it is easy to up-convert to blue or ultraviolet wavelengths needed for nearly any of the popular atomic species in Table \ref{tab:AtomicSpecies}. Typical Ti:S oscillators operate at a repetition rate between $70-100$ MHz, with an output power of about $500$ mW when pumped with a $5$ W green cw laser (typically a frequency doubled Nd-YAG at $532$ nm) \cite{Ye03}. 
       
\begin{figure}[hptb]
\centering
\includegraphics[width=\linewidth,clip]{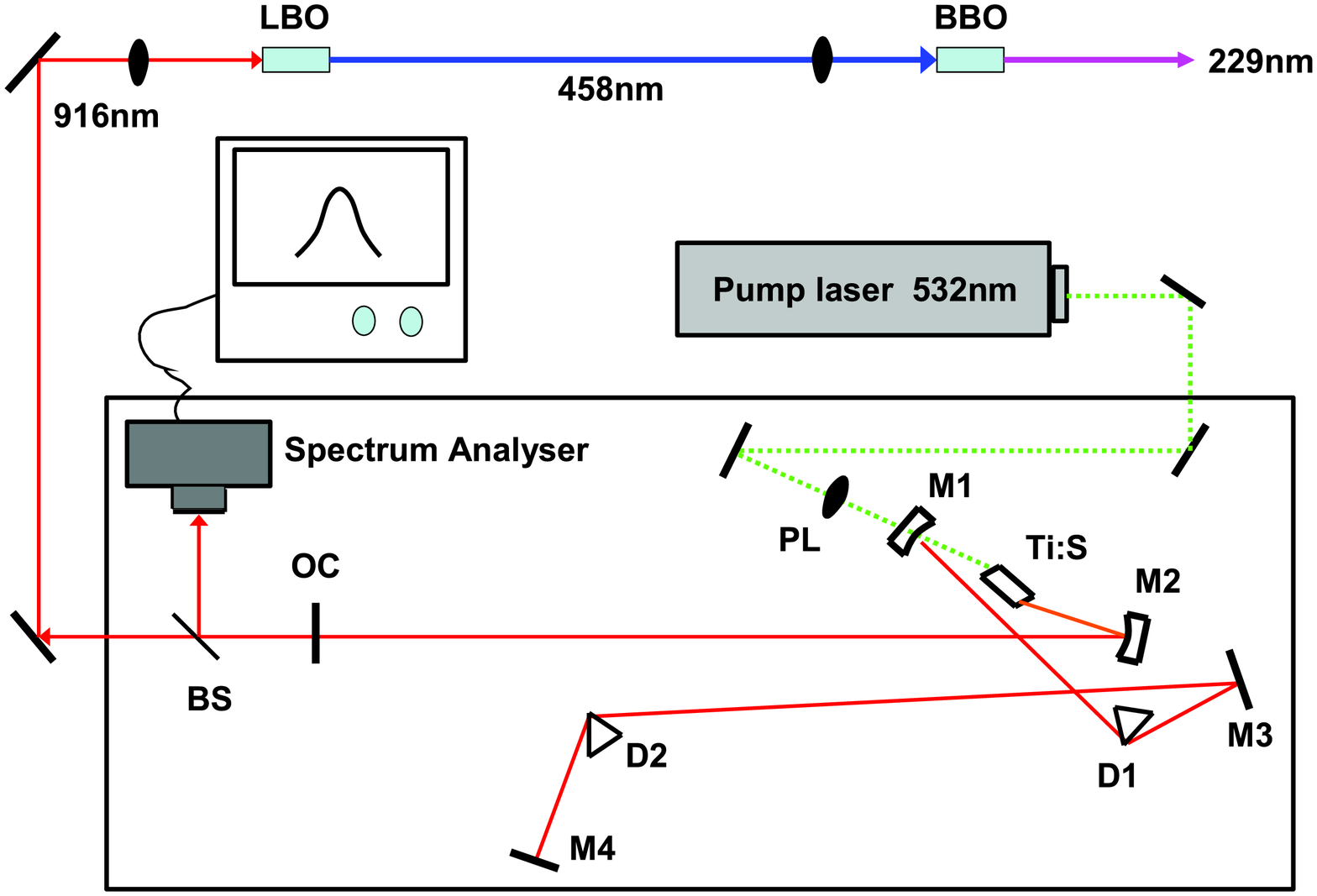}
\caption{Schematic of the cavity for the long-wavelength mode-locked Ti:S pulsed laser. The general configuration is identical to the cavity design of Ref. \cite{Asaki04}. } 
\label{fsecPulsedLaserSetup}
\end{figure}

The long wavelength femtosecond pulsed laser that we built and used in this experiment is similar to a standard femtosecond folded cavity geometry \cite{Asaki04}. However, tuning the cavity to the relatively long central wavelength of $915$ nm requires some important and non-intuitive modifications of the standard setup. Below we give a detailed description of the long-wavelength pulsed laser, as depicted in Fig. \ref{fsecPulsedLaserSetup}. The Ti:S crystal is pumped with typically $5$ W of green light from the output of a Spectra-Physics Millennia and is focused into the crystal with a $12.5$ cm focal-length plano-convex lens (PL). The laser cavity is formed with two flat end mirrors (output coupler OC and high-reflector M4), two highly-reflecting spherical concave mirrors with 20 cm radii of curvature (M1 and M2), and two fused silica prisms used to compensate for the group velocity dispersion (GVD). The asymmetric cavity arrangement gives a total length of 184 cm, with 65 cm in one arm of the cavity, and $119$ cm in the arm containing the dispersion prism pair. The Ti:S crystal from Crystal Systems is $5$ mm long ($3\times 3 \times 5$ mm) and is doped with a gain coefficient of $\alpha=4.44$. The $12.7$ mm diameter output coupler is characterized by a 5$\%$ transmission and a reflective GVD of $-20$ fs$^{2}$ per pass at $915$ nm. This relatively ``closed" output coupler is used to balance the lower gain of the Ti:S crystal when operating at $915$ nm (a transmission of $20\%$ is typically used for ultrafast lasers near $800$ nm). Both $12.7$ mm diameter curved mirrors have $\geq 99.8 \%$ reflection and a reflective GVD of $-50$ fs$^{2}$ at $915$ nm.  The fused silica prism pair are Brewster cut with an apex angle of $69.1^{\circ}$ and give a transmissive GVD of $290$ fs$^{2}$/cm. In order to compensate for the dispersion aquired in passing through the crystal and both prisms, the distance separating the prisms is set to $\sim\!\!~64$ cm. Both $25.4$ mm diameter flat mirrors (M3 and M4) have reflectivities $\geq 99 \%$ and reflective GVD of $-20$ fs$^{2}$. The vendors and part numbers of the laser cavity optics are: OC (LayerTec, GmbH; P/N 101907), M1/M2 (LayerTec, GmbH; P/N 101241), M4 (Newport, Corp, P/N 10B20UF.20), and prism pair (Newport Corp., P/N 06SB10).  
      
%------------------------------FIG---FSEC LASER SETUP----------------
%--[hptb] means p=seperate page, t=top page, b...,h=here
%----------------------------------------------------------------------------  
       
There is an excellent description for this type of cavity when operating a wavelength around $800$ nm in Ref. \cite{Asaki04}. However, when operating at much longer wavelengths, there are a few additional issues. When the laser oscillates at wavelengths longer than about $900$ nm, we observe that heating of the crystal from the pump laser can prevent the laser from mode-locking. We find that by providing good thermal contact between the crystal and its metallic housing and water-cooling the Ti:S crystal assembly eliminates this problem \cite{Hansch}.  In addition, at longer wavelengths, there is a potential problem with the absorption of water vapor in the air, that can lower the gain of the laser to below threshold, especially in the $925-960$ nm range.  

A final critical issue concerns the proper choice of optical coating.  It is known that care must be taken in making sure that cavity mirrors have matched coatings - both curved mirrors are especially sensitive to this requirement \cite{Kapteyn05}. Even if all of the mirrors are from the same coating run, one of them can still prevent proper mode-locking.  This is not well understood, and our experience is that repeated replacement of seemingly identical optical components appears to make a difference \cite{Kapteyn05}. Ultrafast Ti:S lasers operating at any wavelength are sensitive to the subtle details of mirror coatings, but this appears to be even more critical when the laser is tuned to the longer wavelength range where the gain is significantly lower.

\section{Appendix B: Probability of Resonantly-Assisted Photoionization Per Pulse}
In this appendix, we determine the probability of photoionization of an atom illuminated with an ultrafast laser pulse that promotes the atom from a ground state through an intermediate atomic resonance and onto the continuum, following Fig. \ref{EnergyLevels2}. For simplicity, the laser pulse is described by a square wave profile of duration $\tau$ and peak intensity $I$ and is in the weak perturbative regime where no other atomic levels contribute to the photoionization \cite{CohenTannoudji}.

%-----------------------FIG---S-P-E SIMPLE ENERGY LEVELS----------------------------
\begin{figure}[hptb]
\centering
\includegraphics[width=0.4\linewidth,clip]{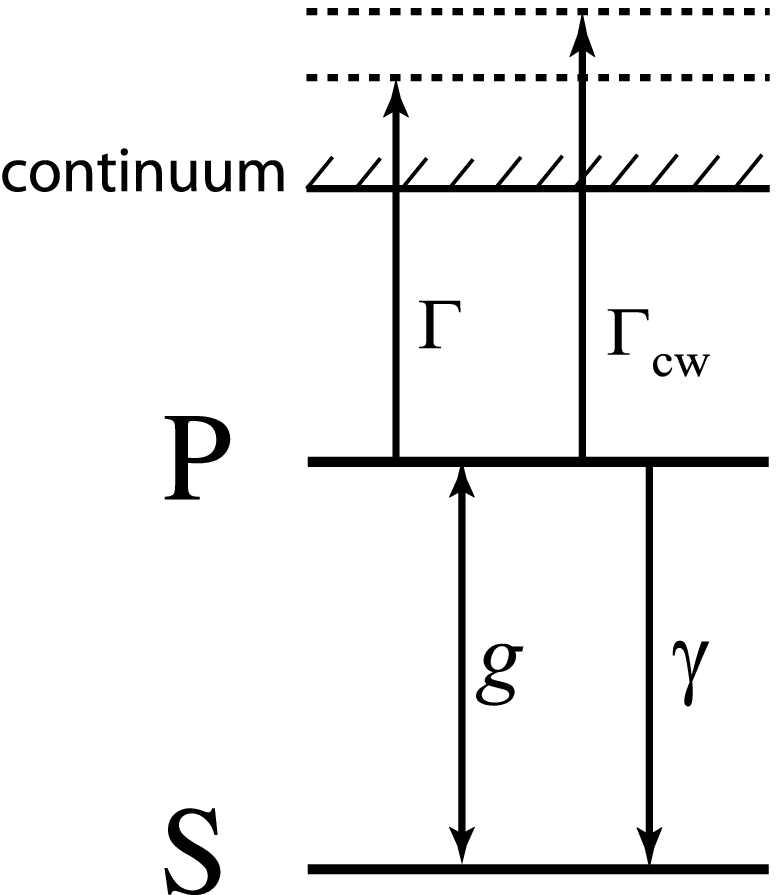}
\caption{Simplified energy level diagram for photoionization from the ground S state through the intermediate P state (radiative decay rate $\gamma$) and into the continuum.  Laser pulses of duration $\tau$ and period $T \gg 1/\gamma$ are resonant with the S-P electric dipole transition, with Rabi frequency $g$ per pulse.  The same laser pulse can ionize the atom from the P state with rate $\Gamma$, or alternatively, a second continuous-wave (cw) laser can ionize the atom from the P state with rate $\Gamma_\text{cw}$.} 
\label{EnergyLevels2}
\end{figure}
%--------------------------------------------------------------------------
We consider an atom initially in the $^1S_0$ electronic ground state coupled to an intermediate exited $^1P_1$ state (radiative linewidth $\gamma$) with Rabi frequency ${g = \gamma\sqrt{I/2I_\text{sat}} \gg \gamma}$, where $I_\text{sat}$ is the saturation intensity for this transition. The Rabi angle accumulated during the pulse is $\theta=g\tau$. The photons in the same laser pulse have enough energy to sequentially ionize the atom with photoionization rate from the ${^1P_1}$ state given by $\Gamma =I\sigma/\hbar\omega$, where $\sigma$ is the ${^1P_1}$ photoionization cross-section and $\omega$ is the laser frequency. In addition, a cw laser with intensity $I_\text{cw}$ used for laser-cooling of the eventual ion can also ionize the neutral atom once in the $^1P_1$ state, with photoionization rate given by ${\Gamma_\text{cw} = I_\text{cw}\sigma_\text{cw}/\hbar\omega_\text{cw} \ll \gamma}$, where the cw laser parameters are defined analogously to that of the pulsed laser.  The Bloch equations describing the populations $\Pi_S,\,\Pi_P$, and $\Pi_\text{ion}$ of the ground $S$ state, intermediate excited $P$ state, and photoionized state, respectively, can be written
\begin{subequations}
\begin{eqnarray}
\frac{d\Pi_S}{dt} &=& -g(t)C + \gamma \Pi_P  \\
\frac{d\Pi_P}{dt} &=& g(t)C - (\Gamma+\Gamma_\text{cw}+\gamma)\Pi_P  \\
\frac{d\Pi_\text{ion}}{dt} &=& (\Gamma+\Gamma_\text{cw})\Pi_P  \\
\frac{dC}{dt} &=& -\frac{g(t)}{2}(\Pi_P - \Pi_S) - \frac{\Gamma}{2}C \, ,
\end{eqnarray}
\end{subequations}
where $C$ is the coherence between the $S$ and $P$ electronic states and the S-P Rabi frequency for the laser pulse is $g(t)=g$ for $0\le~t<\tau$ and $g(t)=0$ for $\tau\le~t\le~T$.
%\begin{equation}
%g(t) = \left\{ \begin{array}{l l}
%  g & \quad \text{for $0\le~t<\tau$}     \\
%  0 & \quad \text{for $\tau\le~t\le~T$}  \\ \end{array} \right.  
%\end{equation}

The $^1P_1$ spontaneous emission lifetime is usually much shorter than the laser pulse period ($T\gg~1/\gamma \gg~\tau$), so the neutral atom returns to the ground $^1S_0$ state before the next pulse (unless it has ionized). The probability $P_\text{ion}$ of ionization per laser pulse period $T$ can then be identified as the value of $\Pi_\text{ion}(t=T)$ by integrating the above Bloch equations over the time span $0\le~t\le~T$, under the initial condition $\Pi_S(t=0)=1$,
%\begin{widetext}
\begin{eqnarray} 
 P_\text{ion} &=& 1-e^{-x}\left(\frac{\theta^2\!\!-\!x^2 \!
       \cos\sqrt{\theta^2\!\!-\!x^2}}{\theta^2\!\!-\!x^2} + \frac{\sin\sqrt{\theta^2\!\!-\!x^2}}{\sqrt{\theta^2\!\!-\!x^2}}\right) \nonumber \\
&+&\frac{e^{-x}}{2}\left(\!\frac{\Gamma_\text{cw}}{\Gamma_\text{cw}+\gamma}\!\right)\!\!
\left(\!\!\frac{\theta^2}{\theta^2\!\!-\!x^2}\!\!\right)\!\!
\left(1-\cos\sqrt{\theta^2\!\!-\!x^2}\right) ,
\end{eqnarray}
%\end{widetext}
where $x = \Gamma \tau\!/2$.  In the weak pulse limit considered here, the S-P Rabi frequency is much larger than the photoionization rate from the P state ($g \gg \Gamma$), thus $\theta \gg x$ and the above equation simplifies to 
\begin{equation}
P_\text{ion} = 1-e^{-\Gamma\tau\!/\!2}\left[1+\frac{\Gamma\tau}{2}\frac{\sin\theta}{\theta}
                    +\frac{\Gamma_\text{cw}}{2\gamma}(1-\cos\theta) \right] .
\end{equation}

\section{Appendix C: The Flux of Photoionized Atoms from a Vapor}
\begin{figure}[hptb]
\centering
\includegraphics[width=0.8\linewidth,clip]{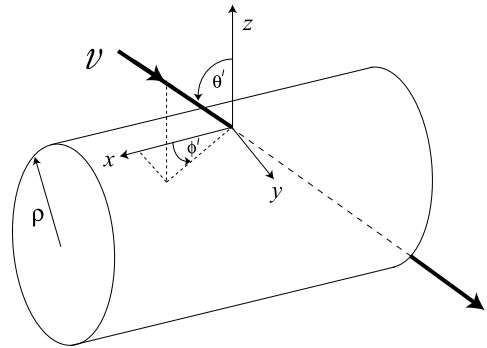}
\caption{Section of cylindrical loading volume of radius $\rho$ and length $L \gg \rho$. An atom moving with speed $v$ enters the surface of the cylindrical volume at polar (azimuthal) angle $\theta'$ ($\phi'$). For $\phi'=0$, the atom crosses the center of the laser beam.} 
\label{Cyl}
\end{figure}

Here, we calculate the average flux of atoms expected to photoionize within the trapping volume.  The photoionizing laser consists of ultrafast pulses separated by period $T$ that are focused to a Gaussian transverse waist of $\rho$.  The trap dimension $L \gg \rho$ is defined as the linear dimension of the trapping volume along the laser beam, and is assumed to be much smaller than the axial Rayleigh range of the laser beam.  The relevant loading volume is thus a cylindrical shape of radius $\rho$ and length $L$ as shown in Figs. \ref{FlyingAtoms} and \ref{Cyl}.
%-----------------------FIG---CYLINDER---------------------------
%--------------------------------------------------------------------------

Given the probability per pulse $P^{(0)}_\text{ion}$ of ionizing a neutral Cd atom at rest in the focus of the Gaussian beam (Eq. \ref{IonizationProb2}), we now consider the probability of photoionization for an atom that pierces the cylindrical loading volume at a radial distance $\rho$ from the laser axis.  The atom has speed $v$, polar (azimuthal) angle $\theta'$ ($\phi'$) with respect to the normal of the cylindrical surface, as drawn in Fig. \ref{Cyl}.  This atom experiences a Gaussian laser beam profile along its trajectory, and the resulting probability of photoionization for a single laser pulse is proportional to the squared intensity,  
\begin{equation}
P_\text{ion}(\theta',\phi') = \tilde{P}^{(0)}_\text{ion}(\theta',\phi')
                                e^{-4\zeta(t)^2\!/\!w(\theta',\phi')^2}\, ,
\end{equation}
where the peak photoionization probability $\tilde{P}^{(0)}_\text{ion}(\theta',\phi')$ and the effective laser beam waist $w(\theta',\phi')$ along the atom's trajectory $\zeta(t)$ are
\begin{equation} 
\tilde{P}^{(0)}_\text{ion}(\theta',\phi')
       = P^{(0)}_\text{ion}\,\,exp\left(-\frac{4\sin^2\theta'\sin^2\phi'}
                                      {1-\sin^2\theta'\cos^2\phi'} \right)
\end{equation}
and
\begin{equation}
w(\theta',\phi') = \frac{\rho}{\sqrt{1-\sin^2\theta' \cos^2\phi'}} \, .
\end{equation}
The net photoionization probability accumulated along the atom's trajectory is then
\begin{equation}
P_\text{net}(v,\theta',\phi') = \\
  1-\!\!\prod_{j=-\infty}^{+\infty}\!\left[1\!-\!\tilde{P}^{(0)}_\text{ion}(\theta',\phi')
    e^{-4\zeta_j^2\!/\!w(\theta',\phi')^2} \right]  ,
\end{equation}
where $\zeta_j = jvT + \zeta_0$ is the position of the atom at the $j$th laser pulse and $\zeta_0$ is the position of the atom for a particular pulse. The above product is safely taken to $\pm\infty$ because the probability of photoionizing an atom is negligible outside of the laser beam where $\zeta_j \gg w(\theta',\phi')$. For $P_\text{ion}(\theta',\phi') \ll  1$, the above product simplifies to
\begin{subequations}
\begin{eqnarray}
P_\text{net}(v,\theta',\phi') 
   &=&  \tilde{P}^{(0)}_\text{ion}(\theta',\phi')\!\!\sum_{j=-\infty}^{+\infty} \!\! 
                               e^{-4\zeta_j^2\!/\!w(\theta',\phi')^2} \;\;\;\;\; \\
   \simeq \tilde{P}^{(0)}_\text{ion}(&\theta'&,\phi')\int_{-\infty}^{+\infty} 
                       \frac{d\zeta}{vT}\,\,e^{-4\zeta^2\!/\!w(\theta',\phi')^2} \\
   = \tilde{P}^{(0)}_\text{ion}(&\theta'&,\phi')\sqrt{\frac{\pi}{4}}\frac{w(\theta',\phi')}{vT} \, .
\end{eqnarray}
\label{integ}
\end{subequations}
The conversion to an integral in Eq. \ref{integ}b presumes that the the atom experiences many laser pulses as it moves through the beam ($vT \ll \rho$).  For thermal Cd atoms of mean speed $\bar{v} \simeq 240$ m/s and a laser repetition rate of $1/T = 80$ MHz, a typical atom moves $\bar{v}T \simeq 3\, \mu$m between pulses, which is much smaller than a typical beam waist.

We can now calculate the flux of photoionized atoms by averaging $P_\text{net}(v,\theta',\phi')\,n(v)v\cos\theta'$ over the thermal density distribution $n(v)$ of atomic speeds $v$ in the hemisphere represented by $\theta' \in (0,\pi/2)$ and $\phi' \in (0,2\pi)$.  The density distribution is 
\begin{equation}
%n(v) = n_0\left(\frac{m}{2\pi~k_BT_v}\right)^{3/2}exp\left(-\frac{mv^2}{2k_BT_v}\right) \, , 
n(v) = \frac{n_0}{(4\pi\bar{v}^2)^{3/2}}\, e^{-\frac{4v^2}{\pi\bar{v}^2}} \, , 
\end{equation}
where $n_0$ is the uniform density of atoms in the vapor, $T_v$ is the vapor temperature, $k_B$ is Boltzmann's constant, and $m$ is the atomic mass \cite{Reif}.  The net flux of ions is then
\begin{widetext}
\begin{subequations}
\begin{eqnarray}
\Phi_\text{ion}\!\! &=&\int_0^{\infty}\!\! v^2 \,dv  
                       \int_0^{2\pi}\!\!\!\! d\phi' \!\! \int_0^{\pi/2}\!\!\sin\theta' d\theta'
                       \, P_\text{net}(v,\theta',\phi') \,n(v) v\cos\theta'  \\
  &=& P^{(0)}_\text{ion}\,\sqrt{\frac{\pi}{4}}\frac{\rho}{T}
        \int_0^{\infty}\!\! v^2 n(v)\,dv
        \int_0^{2\pi}\!\! d\phi'  
        \int_0^{\pi/2}\!\!d\theta'\,\frac{\sin\theta'\cos\theta'}{\sqrt{1-\sin^2\theta' \cos^2\phi'}}   
         \,\,exp\left(-\frac{4\sin^2\theta'\sin^2\phi'}{1-\sin^2\theta'\cos^2\phi'} \right)  \\
%        \!\! &=&\!\! P^{(0)}_\text{ion}\,\sqrt{\frac{\pi}{4}}\frac{n\rho}{T}\left(\frac{1}{4\pi}\right)
%        \int_0^{2\pi}\!\!\!\! d\phi' \!\! \int_0^{\pi/2}\!\!d\theta'\sin\theta'\cos\theta' \,\,   
%         \frac{exp\left(-4\frac{\sin^2\theta'\sin^2\phi'}{1-\sin^2\theta'\cos^2\phi'} \right)}
%               {\sqrt{1-\sin^2\theta' \cos^2\phi'}} \\
         &=& P^{(0)}_\text{ion}\, \left(\frac{n_0\rho}{8T} \right)\, .
\label{IonFlux}
\end{eqnarray}
\end{subequations}
\end{widetext}

\begin{acknowledgments}
We acknowledge useful discussions with J. A. Rabchuk. 
This work is supported by the National Security Agency and the Disruptive Technology Organization under Army Research Office contract W911NF-04-1-0234, and the National Science Foundation Information Technology Research Program.
\end{acknowledgments}

\bibliographystyle{prsty}

\begin{thebibliography}{10}
\bibitem{electrode} When the photoionization laser radiation strikes the trap electrodes, the loading rate is often very high.  This may stem from Cd atoms ablated from the surface and/or electron bombardment loading from electrons released from the electrode surfaces illuminated with the 5.4 eV photons.
\end{thebibliography}

\end{document}